\documentclass[notitlepage,superscriptaddress,10pt,%
floatfix,reprint,prl,aps]{revtex4-1}
\usepackage{graphicx}
\usepackage{amsmath,mathtools}
\usepackage{natbib}
\usepackage[utf8]{inputenc}
\usepackage{amssymb,nicefrac,hyperref}
\usepackage{color}

\usepackage{bbm}

\let \vec \boldsymbol

\newcommand{\leftrarrows}{\mathrel{\raise.75ex\hbox{\oalign{%
  $\scriptstyle\leftarrow$\cr
  \vrule width0pt height.5ex$\hfil\scriptstyle\relbar$\cr}}}}
\newcommand{\lrightarrows}{\mathrel{\raise.75ex\hbox{\oalign{%
  $\scriptstyle\relbar$\hfil\cr
  $\scriptstyle\vrule width0pt height.5ex\smash\rightarrow$\cr}}}}
\newcommand{\Rrelbar}{\mathrel{\raise.75ex\hbox{\oalign{%
  $\scriptstyle\relbar$\cr
  \vrule width0pt height.5ex$\scriptstyle\relbar$}}}}

\makeatletter
\def\leftrightarrowsfill@{\arrowfill@\leftrarrows\Rrelbar\lrightarrows}
\newcommand{\xleftrightarrows}[2][]{\ext@arrow 3399\leftrightarrowsfill@{#1}{#2}}
\makeatother

\begin{document}
\title{Quantitative spectroscopy of single molecule interaction times}
\author{H-H. Boltz}
\email{boltz@zib.de}
\affiliation{Zuse Institute Berlin (ZIB)}
\affiliation{Max Delbrück Center for Molecular Medicine, Berlin}
\author{A. Sirbu}
\affiliation{Max Delbrück Center for Molecular Medicine, Berlin}
\author{N. Stelzer}
\affiliation{Max Delbrück Center for Molecular Medicine, Berlin}
\author{M. J. Lohse}
\affiliation{Max Delbrück Center for Molecular Medicine, Berlin}
\affiliation{University of Würzburg, Institute of Pharmacology and Toxicology}
\author{C. Schütte}
\affiliation{Zuse Institute Berlin (ZIB)}
\affiliation{Freie Universität Berlin, Institut für Mathematik}
\author{P. Annibale}
\email{paolo.annibale@mdc-berlin.de}
\affiliation{Max Delbrück Center for Molecular Medicine, Berlin}

\begin{abstract}
    Single molecule fluorescence tracking provides information at nm-scale and ms-temporal resolution about the dynamics and interaction of individual molecules in a biological environment. While the dynamic behavior of isolated molecules can be characterized well, the quantitative insight is more limited when interactions between two indistinguishable molecules occur. We address here this aspect by developing a solid theoretical foundation for a spectroscopy of interaction times, i.e. the inference of interaction constants from imaging data. The non trivial crossover between a power law to an exponential behavior of the distribution of the interaction times is highlighted here, together with the dependence of the exponential term upon the product of the microscopic reaction rates (affinity). Our approach is validated on simulated as well as experimental datasets.
\end{abstract}
\maketitle

Recent progresses in the field of fluorescence single molecule methods \cite{schmidt1996} have made single molecule tracking (SMT) of membrane receptors in a widefield microscope a technique within reach of many laboratories. This has benefited significantly those researching cell membrane receptor biophysics and pharmacology, since single molecule methods offer the opportunity to probe dynamic processes such as oligomerization~\cite{Calebiro}, interaction with downstream signaling partners~\cite{sungkaworn2017}, trafficking~\cite{Puthenveedu}, as well as conformational changes at the level of the isolated receptor~\cite{Blanchard}. Combinations of efficient and photo-stable labeling strategies with advanced optical imaging methods yield spatial-temporal resolutions that allow for nanometer-level detection~\cite{chenouard2014} and millisecond-temporal resolution tracking of individual molecules~\cite{shen2017}. Despite the apparent relative simplicity of experimental preparation and data acquisition, data analysis and interpretation remain fraught with significant caveats. In particular, given the continuous nature of the molecular point-spread-funcion (PSF), defining the duration of a molecular interaction is not trivial: as two identically labeled particles approach each other, their PSFs become unresolvable. While this issue was addressed in static datasets exploiting the notion of stochastic activation and localization~\cite{Betzig2006}, for dynamic datasets, where all molecules present are visible at the same time, the problem is still present. 
The answer to this question will allow addressing the important problem whether and how these particles are (constructively) interacting, revealing details on microscopic interaction rates.

Previous work~\cite{kusumi, Calebiro, sungkaworn2017, moeller} has addressed this issue by generating and evaluating an histogram (or distribution) of molecular overlap (or colocalization) times.
As the identification of colocalized particles is fundamentally easier, if they are labeled differently, we will only address the (more complex, but also more typical) monochromatic case. Furthermore, we are limited to a density regime, where trimers and higher order clusters are irrelevant and particles are found as either monomers or dimers.

We begin by considering a related problem that can be handled analytically: the distribution $p_{\text{fp}}$ of first-passage times of a freely diffusing particle through a circular boundary. We focus on standard Brownian diffusion, but extension to anomalous diffusion is possible and may be relevant for some systems~\cite{metzler2014}. 

\begin{figure*}[h] 
\includegraphics[width=.9\linewidth]{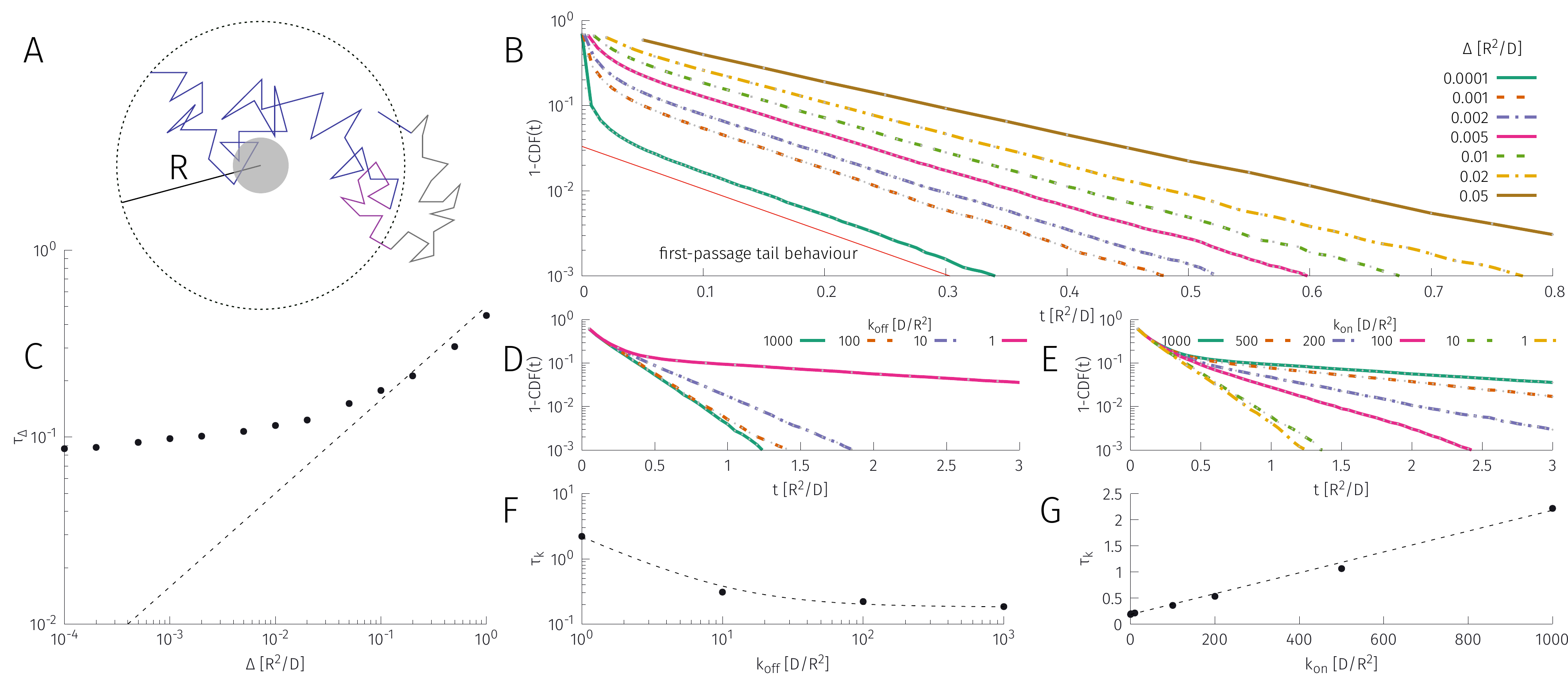}
\caption{A: Sketch of the simulation setup highlighting the difference between the analytically easily accessible first-passage time and the colocalization time at finite exposure time. There are acquisition times for which the particle will be considered always inside the circle, even though the grey part is outside of it. (B): Distribution of colocalization times generated computationally. The family of curves highlights the dependence upon the acquisition time. When the temporal resolution is adequate, the distribution of colocalization times has two functional components: a rapid power law, and a slower exponential decay.  (C): Numerically recovered timescale from the exponential part. The straight line corresponds to $\tau_k \sim \sqrt{\Delta}$ as a guide to the eye. (D) Dependence of the distribution of colocalization times as a function of $k_\text{off}$  and (E) as a function of the $k_\text{on}$. Both use $\Delta =0.05 R^2/D$. The dependence of the resulting slope of the exponential component of the colocalization time distributions in D and E is displayed respectively in F) and G) highlighting the $\tau_k\sim k_\text{off}^{-1}+\text{const}$ and $\tau_k\sim k_\text{on}+\text{const}$ behaviour, respectively.}
\label{fig:coll_times}
\end{figure*}

This will serve as a continuum reference theory to understand the overall shape and dependencies of the overlap time distribution. We denote the particle's diffusion constant by $D$ and the radius of the circle (approximating the PSF) by $R$.
Dimensionally, the only timescale in the problem is $\tau = R^2/D$. Hence, one expects $-\log p_{\text{fp}} \sim t/\tau$ for the asymptotic behavior on the large time tail. For very small times, the particle is blind to the scale of the region and one expects\cite{fisher} a scale-free distribution, $p_{\text{fp}}\sim x^{\alpha}$. This can be summarized as:
\begin{align}
    p_{\text{fp}}(x) &\sim \begin{cases} x^{-\nicefrac{3}{2}} & \text{for } 0<x\ll \tau \\
    \exp{(-t/\tau_{\text{fp}})}  & \text{for } x\gg \tau \end{cases} \label{eq:pt}
    \end{align}
    with $\tau_\text{fp} \approx 1/5.78 \tau$. Details of this calculation have been discussed earlier~\cite{govorun2001,*oshanin2010} and are also provided in the appendix. Similar findings have been made using approaches with explicit model interaction potentials.~\cite{margolin2005} 
    Graphically, these two behaviours can be interpreted as two ``classes'' of trajectories, as summarized in Fig.~\ref{fig:coll_times}: short trajectories (purple) that only explore the rim and long trajectories (blue) that traverse the whole colocalization area/PSF. As we will show below, the overall description provided by eq.~\eqref{eq:pt} holds quite generally also for finite acquisition time and interacting particles.

\begin{figure*}
\includegraphics[width=.9\linewidth]{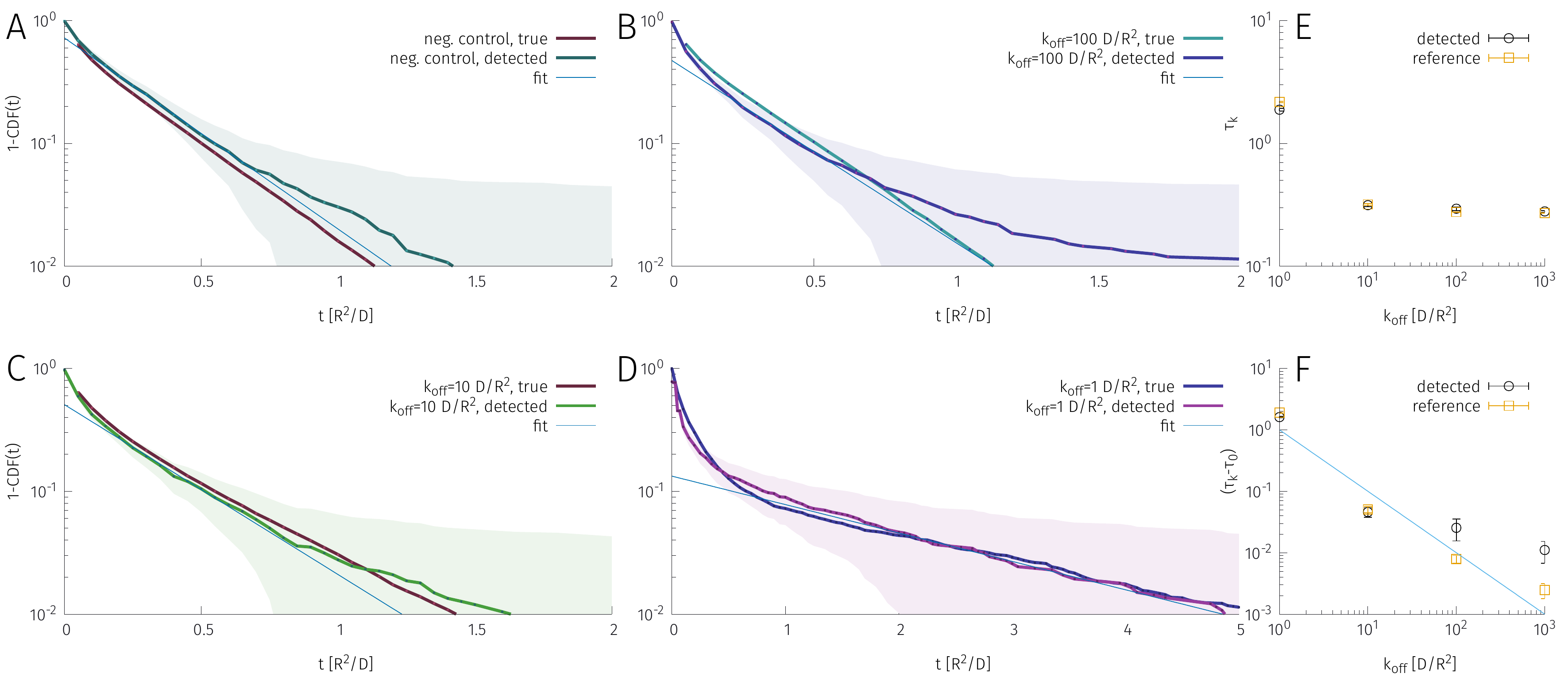}
\caption{Distribution of colocalization times obtained running a commercial single particle tracking software on simulated datasets. (A)-(D) Recovered distribution of colocalization times for a non-interacting control. Displayed are the true interaction values (known from the simulated centers of each molecule at any given time) and the output of the tracking software (detected), as indicated in the legend. The slope of the exponential component of the detected colocalization times distribution is fit to extract the apparent lifetime of the interaction $\tau$. Shaded bands highlight 95\% confidence level for the detected distribution as inferred from the Dvoretzky–Kiefer–Wolfowitz \cite{dvoretzky1956,*massart1990} inequality. (E),(F) Recovered apparent lifetimes  as a function of the interaction strength ($k_\text{off}$). In (F) the background contribution is subtracted.}
\label{fig:utrack_times}
\end{figure*}

In experiments, observation is stroboscopic: images are taken at a finite acquisition rate. Thus, a trajectory that ventures outside the region of interest, cp. the grey segment in Fig.~\ref{fig:coll_times}A, may be considered within the region at every time of observation. The actual problem is to determine the first-{\em observed}-passage when considering snapshots \footnote{We consider images to be taken instantaneously, shifting the effect of a finite exposure time into the PSF.} taken at a fixed acquisition time interval $\Delta$. Thus, we need to determine how the distribution $p_\Delta$ inferred from the discrete acquisition will deviate from the continuous first-passage analysis presented in the previous paragraph. We shall note here that the dimensional arguments still hold, since the first-passage tail timescale $\tau_\text{fp}$ is a lower bound to the observed timescale $\tau_\Delta$ with $p_{\Delta}(x) \sim \exp{(-t/\tau_{\Delta})}$, i.e. $\tau_\Delta \geq \tau_\text{fp}$, as the discrete observation can only extend the apparent colocalization time. Analogously, we expect monotony, $\tau_{\Delta}>\tau_\Delta'$ for $\Delta>\Delta'$. More specifically, we expect $\tau(\Delta) \sim \sqrt{\Delta}$ as the probability within the colocalization region becomes uniformly spread for large times and, thus, the probability to be observed outside grows with $\Delta$ as the size of the zone at the rim that can reach the outside with one step. 

We validated this intuitive argument by numerical simulations of  diffusing particles using direct integration of the Langevin equation, $\mathrm{d}x^n_{i}/\mathrm{d}t = \sqrt{2D}\eta(t)$ with $i=x,y$ indexing spatial dimensions, $n=1..N$ indexing particles and $\eta(t)$ denoting a Gaussian random force with $\langle \eta(t) \rangle=0$ and $\langle \eta(t)\eta(t') \rangle = \delta(t-t')$. The particles are considered colocalized if their distance is not larger than $R$, as depicted in Fig.~\ref{fig:coll_times}B. We show empirical cumulative distributions, $\operatorname{CDF}(t)=1/N_t \sum_i^{N_t} \Theta(t-t_i)$; intuitively one can think of $1-CDF(t)$ as the fraction of colocalization events lasting longer than $t$.

Fig.~\ref{fig:coll_times}B illustrates the time spent inside the circle as a function of the acquisition time. If the acquisition time is adequate \footnote{For practical purposes, a typical order of magnitude for $R^2/D$ is $1$s.} we observe the expected combination of a rapid power law decay and a slower exponential behavior of the distribution. The slope of the exponential part will reflect the presence of an interaction between the particles, if any. 

Interestingly, Fig.~\ref{fig:coll_times}C highlights how the recovered timescale of the exponential tail depends on the acquisition time. As the time $\Delta$ grows higher, the timescale increases, suggesting ``apparent interactions'' also in this negative control. This increase is in line with our expectation $\tau(\Delta)\sim\sqrt{\Delta}$ and really stresses the importance of a careful examination of the behaviour of freely diffusing particles under given imaging conditions.

We then extended this approach to interacting molecules using a Doi-like model \cite{Doi} in which the interaction is described by three parameters: a binding rate $k_\text{on}$, an unbinding rate $k_\text{off}$ and an interaction range $\ell$. Pairs of molecules $i,j$ whose distance is within the range, $r_{ij}<\ell$, bind with constant rate $k_\text{on}$. Rather than introducing an explicit new construct formed this way, we let bound particles continue to diffuse while enforcing a constraint $r_{ij}<\ell$, corresponding to soft molecules which is somewhat justified by noticing that our point particle description neglects most of the internal degrees of freedom. From the bound state, molecule pairs unbind with constant rate $k_\text{off}$ after which they can leave each other's vicinity again.
Previously, the tail behaviour of the overlap time distribution (or the excess time compared to a freely diffusing background) has been identified with $k_\text{off}^{-1}$.  Intuitively, the typical time spent in the bound state should also depend on the binding rate and the range, as those control the probability to be bound and thus the number of unbinding events, each of which takes a time $\sim k_\text{off}^{-1}$.

We recover the expected dependence on $k_\text{off}$, as illustrated in Fig.~\ref{fig:coll_times}D, i.e. the higher $k_\text{off}$ the shorter the lifetime of the interaction. Interestingly, however, also the $k_\text{on}$  plays a role, indicating the importance of recaptures on the observed distribution of interaction times, as shown in Fig.~\ref{fig:coll_times}E.

If we look at the dependence of the recovered lifetime of the interaction $\tau_\text{k}$ (Fig.~\ref{fig:coll_times}F), we observe that  at very high $k_\text{off}$ there is a residual lifetime that agrees with the result for a freely diffusing monomeric particle that is observed also in the negative control (Fig.~\ref{fig:coll_times}A).  Furthermore, as the interaction becomes stronger (smaller $k_\text{off}$), the recovered lifetime scales linearly with $k_\text{off}$ Fig.~\ref{fig:coll_times}F. 
Fig.~\ref{fig:coll_times}G displays the dependence of $\tau_\text{k}$ upon  $k_\text{on}$.

We can formalize our intuition to corroborate these numerical findings (see Appendix), by calculating the typical time spent within the colocalization region as a function of the affinity $K=k_\text{on}k_\text{off}^{-1}$ as
\begin{align}
    \tau(K) &= \tau(0) + \text{const}\cdot K \text{.} \label{eq:tau}
\end{align}

We conclude that introducing an interaction adds a term scaling as $k_\text{off}^{-1}$ to the timescale $\tau(0)=\tau_\Delta$ observed in the exponential tail, but one cannot identify these with each other, in particular there is a dependence on the binding rate $k_\text{on}$. This dependence is relevant in comparative studies where the observed timescales might be used to assess the difference in unbinding rate as a proxy for different dimerization behaviour. Furthermore, the overall gestalt of the overlap time distribution is basically unaltered, i.e. we see the same algebraic initial behaviour crossing over into in an exponential tail with a timescale that follows eq.~\eqref{eq:tau}. The reasoning behind this is that the practically relevant regime is $\ell \ll R$, i.e. the interaction range is small compared to the colocalization range. Thus, the scale-free behaviour, formed by trajectories that explore a small region close to the edge of the colocalization range, will be unaffected by the interaction. Within the graphical approach used earlier (Fig.~\ref{fig:coll_times}A), it is rather intutitive that the overall shape of the overlap time distribution remains unchanged: short trajectories on the rim (purple) corresponding to the algebraic beginning remain unchanged, long trajectories (blue) will eventually hit the interaction region and therefore get extended. As a byproduct, the non-exponential behaviour for short times gets more pronounced when the excess time due to interaction is large.

\begin{figure*}
\includegraphics[width=\linewidth]{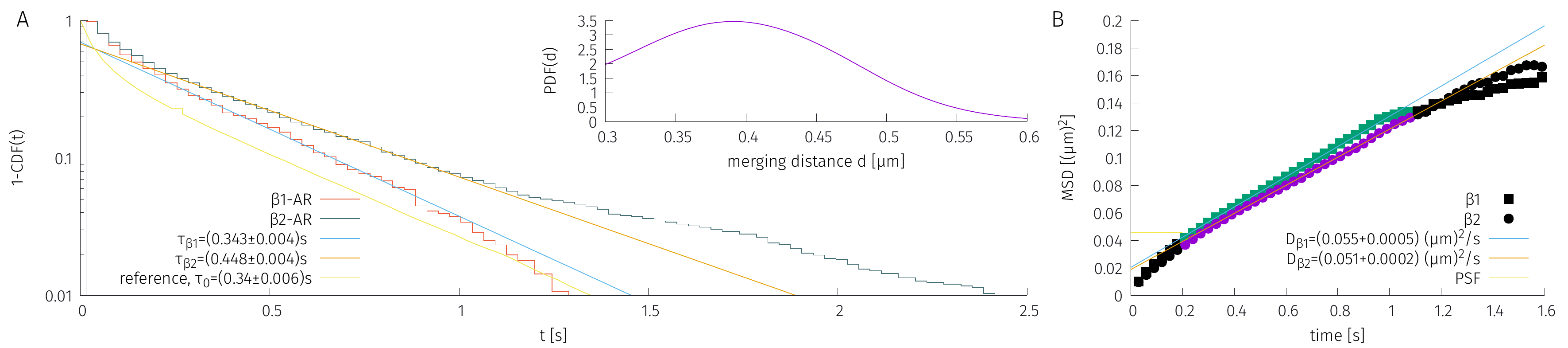}
\caption{Analysis of inter-molecular interaction for the SNAP-$\beta_1$-AR and SNAP-$\beta_2$-AR GPCRs in CHO-cells labeled with SnapSurface 549. (A) Distribution of overlap times for the two $\beta$-AR. Inset: empirical PDF (using Kernel density estimation) of distances of merging particles (larger than the PSF) in the frame before merging. (B) Mean-square displacement as inferred from single-particle tracking for the mutant cells indicating no trivial difference in diffusivities as the origin of the different overlap behavior. The infered values of diffusion constant and threshold distances were used to obtain a suitable reference.}
\label{fig:exp_times}
\end{figure*}

We then asked the question of what results would arise when performing actual single molecule tracking on simulated datasets, mimicking real experimental conditions. We simulated movies \footnote{See Supplemental Material at [URL will be inserted by publisher] for exemplary movies.} of $10^4$ frames, containing $10^2$ particles each, with an acquisition time of $\Delta=0.05 R^2/D$. 
Fig.~\ref{fig:utrack_times} illustrates the results of running an existing, public domain software, u-track \cite{jaqaman2008}, for particle tracking on simulated single molecule movies, where the interaction between the particles is arbitrarily modulated. Fig.~\ref{fig:utrack_times}A-D display the distribution of colocalization times detected by the software on the simulated dataset (green), the ground truth (i.e. the true distribution of colocalization times from the simulated particle positions) and the fit to the detected data in the exponential tail. We generated the ground truth by employing a simple thresholding procedure that identifies every pair with a distance below $R_c$ as colocalized. The detected distribution of colocalization times matches the ``true'' values for the non interacting sample (Fig.~\ref{fig:utrack_times}A), and samples displaying increasing interaction strength (Fig.~\ref{fig:utrack_times}B-D). Once the apparent $k_\text{off}$ (from the fit to the distribution of detected interaction times) is plotted against the simulated $k_\text{off}$, Fig.~\ref{fig:utrack_times}E illustrates that the detected and true values are in excellent agreement. The offset due to the random interaction is clearly visible in the data. When the offset due to random interaction time is subtracted, the recovered $k_\text{off}$ correlates with the simulated one, as displayed in (Fig.~\ref{fig:utrack_times}F).

We then moved to apply this analysis framework to experimental datasets. We chose to analyze the behavior of the two well characterized G-protein-coupled-receptors (GPCRs) $\beta_1$-AR and $\beta_2$-AR.~\cite{Calebiro2013,ValentineHaggie,Annibale2020}
Figure ~\ref{fig:exp_times} illustrates the result of our analysis of single particle TIRF movies of N-terminally SNAP-tagged (a fusion construct allowing extracellular labeling with a bright organic fluorophore of choice) $\beta_1$-AR and $\beta_2$-AR  diffusing on the basolateral membrane of CHO cells. Fig. \ref{fig:exp_times}A displays the detected distribution of colocalization times, together with the distribution of exposure times expected from a randomly diffusing control.
The higher dimer lifetime of the $\beta_2$-AR is in line with previous observations from our group of a higher steady-state dimerization for this receptor \cite{Calebiro}. The lifetime of the $\beta_1$-AR is close to the reference of the non-interacting, reference control.

In this letter, we addressed the question of to what extent a characterization of inter-molecular interactions is possible by means of single-molecule imaging. We have derived the shape of the distribution of overlap times both with and without an interaction  providing a solid theoretical foundation to this important methodology. This analysis reveals that there is a non-exponential, algebraic initial behaviour for small overlap times and an exponential tail for large times, the timescale of which depends on the acquisition time. For interacting molecules, we find that the quantity accessible from the exponential tail (after appropriate correction for the diffusive background) is the affinity, i.e. the ratio of binding and unbinding rates. This would only allow quantitative comparisons between molecules imaged under the same experimental conditions.  These determinations were confirmed when  using a common SMT software on simulated single particle datasets. We finally applied our framework to address the difference in dimerization kinetics between two G-protein coupled receptors, the $\beta_1$- and $\beta_2$-adregeneric receptors.

\begin{acknowledgments} This research is carried out in the framework of the DFG funded Cluster of Excellence EXC 2046 MATH+: The Berlin Mathematics Research Center within the Incubator project IN-B2 to PA, MJL and CS. PA and MJL would like to acknowledge  funding by the Deutsche Forschungsgemeinschaft (DFG, German Research Foundation) through CRC 1423, project number 421152132, subproject C03.
\end{acknowledgments}

\appendix

\subsection{First-passage times}
At some time $t=0$ a point particle subject to Brownian motion enters a radially symmetric (circular/spherical) region of interest (ROI) with radius $R$, we want to determine the distribution $P_l$ of times it lingers within this region, which is the distribution of first-passage time of the distance $r$ of the particle to the center of the ROI.

The Brownian motion is characterized by its Fokker-Planck equation, the diffusion equation,
\begin{align}
\partial_t \Phi(\vec r,t) &= D \Delta \Phi(\vec r,t)
\end{align}
introducing the diffusion constant $D$.

From dimensional analysis, it is obvious that the relevant timescale will be of the form
\begin{align}
\tau &\sim R^2/D
\end{align}
with a corresponding exponential cut-off to the distribution of the lingering times
\begin{align}
- \log P_l^{R,D}(t) &\sim t/\tau \qquad \text{for } t\gg\tau \text{.}
\end{align}

On shorter time-scales, the picture is different. Particles that only linger for a short time effectively do not explore the geometry, eliminating the $R$-scale and making the problem scale-free which would suggest a power-law dependence in $P_l$. Effectively the problem on this time-scale is one-dimensional (the distance to the border) and, thus, we expect to find the one-dimensional return exponent $\chi=3/2$ and, therefore,
\begin{align}
P_l^{R,D}(t) &\sim t^{-3/2} \qquad \text{for } t\ll\tau \text{.}
\end{align}

Due the peculiar notion of the problem with the particle being initially right on the edge of the ROI, special attention is needed to $P_l(t=0)$. Geometrically it is obvious that there is a singular part $P_l(t)\sim 1/2 \delta(t)$ as half of all possibilites for the first motion of a particle right on the edge will lead to the particle not being within the ROI.

To determine the distribution of first-passage times through $r=R$ of a particle that is initially at $r=R-\varepsilon$ ($\varepsilon>0)$, i.e. we are looking for a Green's function $G(\vec r,t)$ to the diffusion equation with the boundary condition $G(\vec r,t)\rvert_{r=R}=0$ and the initial condition $G(\vec r,0) \propto \delta(\vec r-\vec r_0)$ with $r_0=\lvert \vec r_0\rvert=R-\varepsilon$. From this the distribution of passages through the boundary can then be determined from the flux through it
\begin{align}
P_l^\varepsilon(t) = \int_{\lvert\vec r\rvert=R} D \partial_r G(\vec r,t) \text{.}
\end{align}
The approach to determining $G$ is independent of the dimension we use a bilinear decomposition $G(\vec r, t)=\sum_n \exp{\lambda_n t} \psi_n(\vec r) \psi_n(\vec r_0)$ where $\psi_n$ are the eigenfunctions corresponding to
\begin{align}
\lambda_n \partial_t \psi_n &= D \Delta \psi_n
\end{align}
with the stated boundary conditions. 

\paragraph{Three dimensions}
In three dimensions, the radial part of the Laplacian is $\Delta_r \psi_n = \frac{1}{r} \partial_r^2 (r \psi)$ from which we see that a suitable set of solutions is given by $\psi_n = r^{-1}\sin{n\pi r/R}$ with  $\lambda_n = - D (n\pi/R)^2$. Thus, we find
\begin{align}
G(\vec r,t) &\propto\sum_{n=1}^\infty \frac{\sin{\frac{n\pi r}{R}}}{r}\frac{\sin{\frac{n\pi r_0}{R}}}{r_0} \,\mathrm{e}^{- D (n\pi/R)^2 t} 
\end{align}
and
\begin{align}
P_l^\varepsilon(t) &\propto \sum_{n=1}^\infty n \sin {\frac{n\pi \varepsilon}{R}} \,\mathrm{e}^{- D (\pi n/R)^2 t} \text{.}
\end{align}
From this, we can read the dominant timescale, $\tau_1= R^2/(\pi^2 D) \approx 1/(9.87) R^2/D$.

\paragraph{Two dimensions}
In two dimensions, things are fairly similar. We have to replace the sine by the zeroth Bessel function $J_0$ (with $J_1=-J_0'$ denoting the first Bessel function) of the first kind and whose zeroes $\alpha_n$ determine the relevant modes, yielding
\begin{align}
G(\vec r,t) &\propto  \sum_{n=1}^\infty J_0(\frac{\alpha_n r}{R}) J_0(\frac{\alpha_n r_0}{R}) \mathrm{e}^{- D (\alpha_n/R)^2 t} 
\end{align}
and
 \begin{align}
\begin{split}P_l^\varepsilon(t) &\propto \sum_{n=1}^\infty - \alpha_n J_1(\alpha_n) J_0(\alpha_n r_0/R)  \mathrm{e}^{- D (\alpha_n/R)^2 t} \text{.} \end{split}
\end{align}
There is ample room for simplification, but the important part is that the dominant (longest) timescale is given by $\tau_1=R^2/(\alpha_1^2 D) \approx 1/5.78 R^2/D$.

For a more general walk with $\Delta x^2 \sim  \Delta t^{2 \mu}$ ($\mu=1/2$ being the standard Brownian case), a full analytical treatment might not be possible, but we can easily extend the intuitive arguments, by noting that the general return exponent would then be $\chi=1+\mu$. Thus, a careful analysis of the initial behaviour of the overlap time distribution (if available due to time resolution) can provide insight into the (short-time) dynamics.

Finally, we note that the inital $t^{-3/2}$ behaviour for standard Brownian motion is indeed independent of dimension. This comes from the fact that the very fast trajectories effectively only explore one dimension (away from the boundary and back). However, the region of times on which this will be observable is dimensionally dependent (the algebraic part being shorter in higher dimensions).

\subsection{Estimation of effective timescale}
We are interested in the typical time two particles spent colocalized that experience an attractive, binding interaction that is characterized by some rates $k_\text{on},k_\text{off}$ and a range that is small compared to the colocalization region. We can make some
 ground by considering a very rough description of the system formed by one pair by means of four states with Markovian transitions\cite{WS2016}: 1) $\mathsf{A}$, the pair distance is within the colocalization region, but outside the interaction range; 2) $\mathsf{B}$, the pair distance is within the interaction range, but the pair is not bound; 3) $\mathsf{B'}$, the pair is bound; 4) $\mathsf{C}$, the pair distance is outside the colocalization region. For the overlap time distribution, we are interested in the typical timescale the system takes from state $A$ to state $C$. We describe the dynamics by introducing rates $k_\text{off},k_\text{on},k_{AB},k_{BA},k_{AC}$ that control the transitions 
\begin{align}
   \mathsf  B' \xleftrightarrows [k_\text{off}]{k_\text{on}} \mathsf B \xleftrightarrows[k_{BA}]{k_{AB}} \mathsf A \xrightarrow[]{k_{AC}} \mathsf C \text{.} \label{eq:markov}
\end{align}

To keep things simple, we first consider only the subsystem formed by
\begin{align*}
    \mathsf  B' \xleftrightarrows [k_\text{off}]{k_\text{on}} \mathsf B \xrightarrow[]{k_{BA}} \mathsf A
\end{align*}
and ask for the time it takes from state $\mathsf B$ to $\mathsf A$ for the first time. Generally, speaking this will consist of cycles $\mathsf B, \mathsf B', \mathsf B$ followed by single transition to $\mathsf A$. The probability to transition from $\mathsf B$ to $\mathsf B'$ is given by
\begin{align*}
    P_{BB'} &= \int_0^\infty\mathrm{d}t_1 k_\text{on} \mathrm{e}^{-k_\text{on} t_1} \int_{t_1}^{\infty} \mathrm{d}t_2 k_{BA} \mathrm{e}^{-k_2t_2} \text{.} 
\end{align*}
This is a mere reflection of the logic that is at play in the Gillespie algorithm: both possible transitions have exponentially distributed transition times and the one with the smaller time actually happens. Of course, this is readily evaluated to
\begin{align*}
    P_{BB'}=\frac{k_\text{on}}{k_\text{on}+k_{BA}} \text{,}
\end{align*}
from which we also directly gather that the expected number of trials for the first occurrence of $\mathsf A$ is
\begin{align}
    n&=\frac{k_\text{on}+k_{BA}}{k_{BA}}\text{.}
\end{align}
The average time spent on this is given by
\begin{align}
    \tau_{BA} &=\frac{n}{k_\text{on}+k_{BA}}+\frac{n-1}{k_\text{off}} \\
    &= \frac{1}{k_{BA}}+\frac{k_\text{on}}{k_\text{off}k_{BA}}
\end{align}
which corresponds to an effective rate
\begin{align}
    \tilde{k}_{BA}&=\frac{1}{k_{BA}^{-1}+\frac{k_\text{on}}{k_\text{off}k_{BA}}} \text{.}
\end{align}

We can now address the effective rate $k_\text{eff}$ for the transition between $\mathsf{A}$ and $\mathsf{C}$ by doing the same steps mutatis mutandis for the system
\begin{align*}
    \mathsf  B \xleftrightarrows [\tilde k]{k_\text{AB}} \mathsf A \xrightarrow[]{k_{AC}} \mathsf C
\end{align*}
and using $K=k_\text{on}/k_\text{off}$ to end up with
\begin{align}
    k_\text{eff} &= \frac{1}{k_{AC}^{-1}+\frac{k_{AB}}{\tilde{k} k_{AC}}} \\
    &= \frac{k_{AC}}{1+\frac{k_{AB}}{k_{BA}}(1+K)}
\end{align}
which reproduces the result given in eq.~\eqref{eq:tau} for the typical timescale. The realm of applicability of this expression is not limited to the validity of eq.~\eqref{eq:markov} as also non-markovian transition rates (as would be applicable for small times, see discussion in main text) would ultimately lead to this result.  Another approach to the same result, is the reasoning that the effective time should only depend on the probability to be bound which is controlled by the affinity and, thus, would generally lead to a linear term in leading order.

In principle, a very small binding rate $k_\text{on}$ (or, similarly, a small interaction range) could lead to a crossover in the tail from $\tau(0)$ to $\tau(K)$. However, having a small likelihood for the interaction to take place in the first place would inevitably make it hard to observe. Additionally, this in turn would require a really small unbinding rate, so that the two timescales differ significantly. In summation, we expect eq.~\eqref{eq:tau} to be an adequate description of the tail behaviour that is actually observed experimentally.
\end{document}